
\overfullrule=0pt
\magnification=1000
\vsize=21.0 truecm
\hsize=14.5 truecm
\baselineskip=12pt
\hoffset=1.5 truecm

\font\ftext=cmr10
\font\ftit=cmbx10
\font\abstract=cmr10

\parskip=6pt
\parindent=2pc

\font\titulo=cmbx10 scaled\magstep1

\def\ii{\'\i}

\def\section#1{\vskip 0.5truepc plus 0.1truepc minus 0.1truepc
	\goodbreak \leftline{\titulo#1} \nobreak \vskip 0.1truepc
	\indent}
\def\frc#1#2{\leavevmode\kern.1em
	\raise.5ex\hbox{\the\scriptfont0 $ #1 $}\kern-.1em
	/\kern-.15em\lower.25ex\hbox{\the\scriptfont0 $ #2 $}}

\def\Real{{\rm I\!R}}  

\rightline{ICN-UNAM, Mexico, January 20, 1994.}

\vskip 1.5pc

\centerline{\ftit HOMOTOPY INVARIANTS AND TIME EVOLUTION}

\vskip 0.3pc

\centerline{\ftit IN (2+1)-DIMENSIONAL GRAVITY  \footnote{*}{
\ftext{This work is supported in part by the National Science
Foundation, Grant No. PHY89-04035, by CONACyT grant No.
400349-5-1714E and by the Association G\'en\'erale pour la
Coop\'eration et le D\'eveloppement (Belgium).}}}

\vskip 1.0pc

\centerline{H. Waelbroeck and F. Zertuche ${}^{\dagger}$ }

\vskip 0.5pc

\centerline{Institute for Theoretical Physics, UCSB}
\centerline{Santa Barbara, CA 93106-4030}
\centerline{and}
\centerline{Instituto de Ciencias Nucleares, UNAM}
\centerline{Apdo. Postal 70-543, M\'exico, D.F., 04510 M\'exico.}

\vskip 0.5pc

\centerline{${}^{\dagger}$Instituto de Investigaciones en}
\centerline{Matem\'aticas Aplicadas y en Sistemas, UNAM,}
\centerline{Secci\'on Cuernavaca, A.P. 139-B}
\centerline{62191 Cuernavaca, Morelos, Mexico}

\vskip 1.0pc
\abstract

\noindent Abstract:  We establish the relation between
the $ISO(2,1)$ homotopy invariants, and the polygon representation
of $(2+1)$-dimensional gravity. The polygon closure conditions,
together with the $SO(2,1)$ cycle conditions, are equivalent to
the $ISO(2,1)$ cycle conditions for the representations
$\rho: \pi_1(\Sigma_{g,N}) \to ISO(2,1)$. Also, the symplectic
structure on the space of invariants is closely related to that
of the polygon representation. We choose one of the polygon
variables as internal time and compute the Hamiltonian, then
perform the Hamilton-Jacobi transformation explicitly.  We make
contact with other authors' results for $g=1$ and $g = 2$
($N = 0$).


\vfill
\eject

\section{1. Introduction}

 This article is the first of a set of two, which pursues the
ultimate goal of finding non-perturbative solutions of quantum
gravity in $2+1$ spacetime dimensions.  The classical theory
will be reviewed here, with an emphasis on the issue of time
and dynamical evolution.

	In one approach to quantum gravity, one focuses on finding a
complete set of gauge invariants, or ``observables", which span
the so-called ``reduced phase space".  One then postulates that
the wave function is a square integrable function on the reduced
configuration space and promotes the canonical brackets to
commutators.  Since the ``gauge symmetries" of gravity include
translations in time, all observables must be constants of the
motion. This leads to one aspect of the problem of time in quantum
gravity:  How does one formulate dynamical questions in such a
reduced phase space and, do the answers to such questions depend
on the choice of time which one makes in order to formulate them?

	Following Witten's work [1], many authors have recently turned
to the toy model of $(2+1)$-dimensional gravity to investigate this
and other longstanding problems of quantum gravity in a simpler
setting [2]. Witten found that the spacetimes with topology
$\Sigma_g \times [0,1]$ can be labeled by representations of
the fundamental group $\pi_1(\Sigma_g) \to ISO(2,1)$.
The $ISO(2,1)$ scalars derived from elements of such a
representation, or homotopy invariants, form an over-complete
set of phase space observables, which are related by non-linear
constraints.  To complete Witten's programme at the classical
level, one must solve these constraints to identify the reduced
phase space explicitly, and solve the dynamical problem.

 The task of solving the constraints and deriving
the reduced phase space was accomplished by Urrutia and one of us
(F.Z.) for genus one [3], based on the work of Nelson, Regge and
F.Z. [4].  Nelson and Regge have since found the reduced phase
space in the case of $SO(2,2)$ for genus 2 [5], and implicitly
for $g > 2$ [6].

	Although the homotopy invariants provide a complete phase space
with only constants of the motion, that does not mean that there
is no dynamics in $(2+1)$-dimensional gravity: as Moncrief
has emphasized [7], the observables should be interpreted as
Hamilton-Jacobi variables [8], which are related to the initial
data for the dynamical system.  Moncrief showed that the homotopy
invariants can be written in relation to the ADM variables and
the extrinsic curvature time $K=t$ [9].  By inverting this relation,
one obtains the ADM variables as functions of the homotopy
invariants and time.  He carried out this procedure in the case
of genus one, thereby solving the time evolution problem in that
case.  Moncrief's work for $g=1$ was carried over to the quantum
theory by Carlip [10] and Anderson [11], with different orderings
leading to different quantum theories.

	To pursue this program beyond genus one, we turn to a different
formulation of $(2+1)$-dimensional gravity.  The explicit solution
of the time-evolution problem for genus g and N particles was
given by one of us (HW) in terms of time-dependent global
variables, which define a polygon representation of spacetime.
Since these variables evolve in time, they are not observable
in the same sense as the homotopy invariants, yet they are what
one would intuitively call ``observable", in the sense that they
can be measured by an observer [12].  Carlip [13] has suggested
that the variables of the polygon representation are
related to the $ISO(2,1)$ homotopy invariants, yet it has not
been clear, until now, exactly how.

	In this article, we will construct a representation of
$\pi_1(\Sigma_g) \to ISO(2,1)$ explicitly as a function of the
polygon variables.  We will then choose an internal time,
compute the Hamiltonian and perform the Hamilton-Jacobi
transformation. This work is prerequisite
to the discussion of quantum dynamics in (2+1)- dimensional gravity,
which we leave to the following article.

	The next two sections are devoted to a review of the Witten
formalism, and of the polygon representation.  The two formalisms
are compared in Sec. 4.  We choose an internal time in Sec. 5,
compute the corresponding Hamiltonian, then perform the
Hamilton-Jacobi transformation in Sec. 6.  Two examples (genus one
and genus two) are worked out in Sec. 7, where we make contact
with the work of other authors.

\vfill
\eject

\section{2. ISO(2,1) Homotopy Invariants}

	Three-dimensional spacetimes are flat everywhere the source
term vanishes, as a consequence of Einstein's equations [14]:
any open ball with trivial topology and no sources is isomorphic
to an open ball in Minkowski space.  The spacetime itself is
isomorphic to a subset of (2+1)-dimensional Minkowski space,
with boundary points identified by elements of a subgroup $\Gamma
\subset ISO(2,1)$. This subset is a cylinder with polygonal base,
which is the basis of the ``polygon representation".  Since $(2+1)$-
dimensional spacetimes are flat, the holonomy maps $\pi_1(\Sigma_{g,N}
)$ into a subgroup of $ISO(2,1)$: this subgroup is precisely $\Gamma$,
as we will see below.  For their relation to the first homotopy
group, we will refer to elements of $\Gamma$ as ``$ISO(2,1)$ homotopies"
in what follows.  Superspace is the space of these subgroups
$\Gamma$ modulo $ISO(2,1)$ conjugacy; for a genus g surface with
N=0 punctures it is locally isomorphic to the $(12g-12)$-dimensional
cotangent bundle of Teichmuller space [15].

	We restrict our discussion in this section to $N = 0$, for simplicity;
the extension to $N \not= 0$ will be considered later on.  The
fundamental group $\pi_1(\Sigma_{g})$ is finite-generated:  One
can choose a basis of $2g$ loops $\{ u_i, v_i, i=1,...,g \}$ of
$\pi_1(\Sigma_{g})$, where each $u_i$ intersects $v_i$ once
(Fig. 2.1). We will use the characters $M_i$ and $E_i$, respectively,
for the $SO(2,1)$ and $T(2,1)$ projections of the homotopies
$\rho (u_i)$:

$$\rho (u_i) = \pmatrix{
M_i & E_i \cr
0 & 1 \cr}  \eqno(2.1)$$

	There is a symplectic structure on the space of ISO(2,1) homotopies,
inherited from the brackets of the $ISO(2,1)$ connection field,
which derive from the Chern-Simons action [1,16].  The equivalence
of sets of homotopies by $ISO(2,1)$ conjugacy represents a
global Poincar\' e symmetry which is generated, in the Hamiltonian
formulation, by six first-class constraints: the so-called ``cycle
conditions" (Eqn. (4.13) below).  Specifically, the global Poincar\' e
transformations are the usual Lorentz frame transformations,
together with the translations of the base-point $O$ of the
homotopy group.  If one thinks of this point as the ``observer",
then timelike translations of $O$ should be interpreted as
``time evolution".  Observables are Poincar\' e invariants and can
be constructed by taking any products of homotopies $\rho (u_i)
\subset \Gamma$ and constructing the two invariants $L_1$ and
$L_2$ as follows [17], [3].

$$L_1 = tr (M) \eqno(2.2)$$

$$L_2 = P \cdot E\eqno(2.3)$$

where

$$P^a = {1\over 2} \epsilon^{abc} M_{cb} \eqno(2.4)$$

	These invariants, for all loops in $\pi_1(\Sigma_{g})$, are
Witten's observables.  They are clearly not all independent,
since the fundamental group of $\Sigma_g$ is finite generated.
As Nelson and Regge showed, one can construct a finite but still
redundant set of invariants which are related by given
non-linear constraints. One then solves these constraints and
finds an ideal of the constraint algebra; the symplectic
structure on this ideal is inherited from the symplectic
structure on the space of $ISO(2,1)$ matrices, leading to a
true reduced phase space.  This program is difficult to carry
out explicitly, but has been completed for genus two by Nelson
and Regge [5].  The group $ISO(2,1)$ can be replaced in all
the above by $SO(3,1)$, $SO(2,2)$ and their supersymmetric
generalizations [4], [18]; the reduced phase space differs
in each case.

	In this approach, one is solving all of the constraints
explicitly, and seeking a reduced phase space spanned by
variables that are invariant under the $ISO(2,1)$ transformations,
which include time translations. How can one formulate
dynamical questions in this picture?  To see what has happened,
we will consider the elementary example of a free
non-relativistic particle, in parametrized form:

$$\{ P_i, X_j \} = \delta_{ij} \eqno(2.5)$$

$$\{ P_T, T \} = 1 \eqno(2.6)$$

$$P_T + {p^2\over 2m} \approx 0 \eqno(2.7)$$

	The Hamiltonian is a linear combination of the constraint
with an arbitrary parameter, which determines the ``time"
parametrization, and can be any function on phase space,
$\lambda (x_i, p_i, T, P_T; t)$.  The ``weak equality" ($\approx $)
used for the constraint reminds one that the equality
cannot be used inside a Poisson bracket, since its bracket
with other variables is not zero (it is the generator of
time translations) [19].

$$H \equiv \lambda (P_T + {p^2 \over 2m} ) \approx 0 \eqno(2.8)$$

$${d T \over dt} = \{ H, T \} \approx \lambda \eqno(2.9)$$

$${dx_i \over dt} = \{ H, x_i \} \approx \lambda {p_i \over m}
\eqno(2.10)$$

$${dp_i \over dt} \approx 0 \eqno(2.11)$$

Combining (2.9) and (2.10) gives the more familiar result

$${dx_i \over dT} \approx {p_i\over m} \eqno(2.12)$$

	Following the Witten-Nelson-Regge approach described above,
one would construct a maximum set of independent ``gauge"
invariants.  The dimension of the reduced phase space is
equal to the number of variables minus the number of
constraints, minus the number of symmetries, in this case
six for $x_i$ and $p_i$, plus two for $T$ and $P_T$, minus two
for the Hamiltonian constraint and the corresponding symmetry.
Any set of six independent invariants can be chosen; a
simple choice could be the following:

$$P_i = p_i \eqno(2.13)$$

$$X_i = x_i - {p_i \over m} T  \eqno(2.14)$$

\noindent The reduced system inherits the following brackets
and Hamiltonian

$$\{ X_i, P_j \} = \delta_{ij} \eqno(2.15)$$

$$H = 0 \eqno(2.16)$$

\noindent All observables are constants of the motion.

	Of course (2.13)-(2.16) is just the Hamilton-Jacobi formalism
for a free particle.  $X_i$ is the initial value of the dynamical
variable $x_i$; the time evolution is given by the way in which T
appears explicitly in the relation (1.14), namely by inverting it:

$$x_i (T) = X_i + {p_i\over m} T \eqno(2.17)$$

Thus, although the variables $X_i, P_i$ are a complete set of
labels for the trajectories, knowing these variables is not
sufficient to answer dynamical questions.

We pause here to comment on the difference between ``initial
conditions" and ``dynamical invariants". While initial conditions
can only be measured on the initial slice, dynamical invariants can
be measured anywhere along the history of the system.  In this sense,
the knowledge of dynamical invariants (energy, angular momentum, etc.)
provides important information about the dynamics.  Since the homotopy
invariants are independent of the base-point $O$ of $\pi_1 (\Sigma_g)$,
they are not attached to any ``initial surface" and one would naturally
think that they are dynamical invariants, rather than initial conditions.
Having found as many invariants
as there are phase space dimensions, one would conclude that
$(2+1)$-dimensional gravity is static. The reason why this is not
the case, is that in the parametrized formalism ``time" is one of the
dynamical variables which can be measured anywhere along the
trajectory of the system, therefore initial conditions, such as
$x_i(T) - (p_i/m) T$, can be measured by observers away from the initial
slice. So the difference between ``initial conditions" and ``dynamical
invariants" does not exist in the parametrized formalism, until one has
specified a choice of internal time. We will see later that when ``time"
is taken to be the size of a loop of $\pi_1(\Sigma_g)$, some of the
invariants appear as initial conditions, much as $X_i(T)$ in Eqn. (2.14),
while others are independent of $T$ and therefore are dynamical
invariants. Which observables are initial conditions, and which are
dynamical invariants, depends on the choice of internal time. Since
$(2+1)$-dimensional gravity is dynamical, there cannot be an internal
time where all of the homotopy invariants are dynamical invariants.
Nevertheless, the knowledge of a complete set of invariants for
$(2+1)$-dimensional gravity has proven to be extremely valuable in
understanding the physical content of the theory.

\vfill
\eject

\noindent
\section{3. The Polygon Representation}

	The solution of the time-evolution problem of $(2+1)$-
dimensional gravity, for spacetimes with the topology
$\Sigma_{g, N} \times (0,1)$ was originally derived [12] from
an exact lattice theory [20].  This solution gives a representation
of spacetime by means of a polygon embedded in Minkowski
space, with identifications of boundary points by elements of
$\Gamma \subset ISO(2,1)$.  A one-to-one map between this
set of polygons and the flat spacetimes was given in [21].
We will first review the constrained Hamiltonian system, then
explain how it is related to $(2+1)$-dimensional gravity.
Later in this article, we will show that the polygon
representation is closely related to the $ISO(2,1)$ Chern-
Simons theory, in particular, the Poisson brackets of the
homotopy invariants will be recovered from the brackets
(3.1) and (3.2).

 The phase space variables are three-vectors $E(\mu), \mu =1,...,2g+N$,
and Lorentz matrices in $2+1$ dimensions, $M(\mu)$, with the
following Poisson brackets.

$$\{ E^a (\mu), E^b (\mu) \} = \epsilon^{abc} E_c (\mu) \eqno(3.1)$$

$$\{ E^a(\mu), M^b{}_c(\mu) \} = \epsilon^{abd} M_{dc}(\mu) \eqno(3.2)$$

\noindent All other brackets are zero,
in particular any brackets of variables with different loop
indices ($\nu \not= \mu$) vanish, and so do all brackets
of M's among themselves.  The dynamics and gauge symmetries
are generated by six first-class constraints $J^a \approx 0$ and
$P^a \approx 0$, which have a Poincar\' e algebra with the
brackets given above.

$$J \equiv \sum_\mu (I - M^{-1}(\mu)) E(\mu) \approx 0 \eqno(3.3)$$

$$P^a \equiv {1\over 2} \epsilon^{abc} W_{cb} \approx 0 \eqno(3.4)$$

\noindent where

$$\eqalign{& W = \biggl( M(1) M^{-1}(2) M^{-1}(1) M(2)\biggr)
\cdots \biggl( M(2g-1) M^{-1}(2g) \cr
& M^{-1}(2g-1) M(2g)\biggr) M(2g+1) M(2g+2) \cdots
M(2g+N)\cr} \eqno(3.5)$$

\noindent and the constraints which define the rest-mass
of each point source,

$$H(\mu) \equiv P^2(\mu) + sin^2 \left( \Omega (\mu)\right) \approx 0
\eqno(3.6)$$

\noindent $( \Omega(\mu ) = 4\pi G m(\mu ); \ \mu=2g+1, \cdots , 2g+N )$.
The condition $W \approx I$, related
to the ``cycle condition" of Riemann surfaces, must be satisfied
by the generators of any representation of the fundamental group,
$\pi_1(\Sigma_g)$.
We will consider only faithful representations in $SO(2,1)$. In this
way, we select among the various sheets of solutions
to (3.5) the one with maximal Euler class, which gives
physically acceptable spacetimes: solutions on other
sheets represent spacetimes with large ``negative mass"
singularities, $\Omega_n = \Omega - 4 \pi n/G$ [1], [12].

	The first constraint has an important geometrical interpretation:
The vectors $E(\mu)$, and their images $M^{-1}(\mu)E(\mu)$ under
the corresponding $SO(2,1)$ matrices, form a closed polygonal
contour in Minkowski space.  Because of this geometrical
interpretation, we refer to the constrained system (3.1)-
(3.6) as the ``polygon representation".  We will require that
the three-vectors $E(\mu)$ and all diagonals of the polygon are
spacelike; this is necessary and sufficient for the spacetime to
admit a spacelike foliation locally in time [21] (it cannot
admit a spacelike foliation for all t [22]).

	The Hamiltonian for this parametrized system is a linear
combination of the constraints.  The constraints $J^a \approx 0$
generate Lorentz transformations; the constraints $P^a \approx 0$
are the energy-momentum constraints and the corresponding
parameters in the Hamiltonian are the ``lapse-shift functions";
their geometrical interpretation will be made clear shortly.
If one chooses a fixed frame, the Hamiltonian constraint becomes

$$H \equiv \sum_a N_a P^a + \sum_\mu N(\mu) H(\mu) \approx 0
\eqno(3.7)$$

	The matrices $M(\mu)$ are constants of the motion, since H
depends only on M's and brackets among M's are zero; on the
other hand, the time derivative of a three-vector $E(\mu)$ is
a function of M's, and therefore a constant of the motion.  Thus,

$${dM(\mu) \over dt} = 0 \eqno(3.8)$$

$${d^2E(\mu)\over dt^2} = 0 \eqno(3.9)$$

	These variables represent a spacetime which we will construct in
a moment.  First note that the edges of the polygon (3.3) come in
pairs, with the elements of a pair identified by the corresponding
matrix (for example, $E(1)$ and $- M^{-1}(1)E(1)$).  Let N be a
timelike three-vector at one corner of the polygon.  This corner
belongs to two edges, each of which is identified to another one.
Following these identifications, the corner is mapped to two other
corners and N to two timelike vectors (such as $M^{-1}(1) N$).
These corners and timelike vectors each belong to one other edge,
which is identified to another edge of the polygon, etc.  In this
way each corner of the polygon is endowed with a timelike three-
vector, which is the image of N under the corresponding
identification.  This procedure terminates when an identification
takes one back to a corner of the polygon which had already been
endowed with a timelike vector.  This last identification returns
the same timelike three-vector as was already there, if and only
if the $SO(2,1)$ cycle condition (3.4) holds.  A corner which
respresents the location of a point source is not identified to
any other, but lies at the intersection of two identified edges.
The Lorentz matrix which identifies these two edges is elliptic
(a rotation); the axis of this rotation is proportional to the
energy-momentum three-vector of the particle and the angle is
related to its mass.  We place a timelike line parallel to this
axis at each such corner.  We will show next how to construct
the spacetime from the initial polygon and these timelike lines.

	For concreteness, consider the example of genus one.  The
polygon is a quadrilatteral figure in Minkowski space, $ABCD$
[Figure 3.1].  The edge $AB$, represented by the three-vector
$E(1)$, is mapped to $DC$ by the matrix $M^{-1}(1)$, and $BC$ to
$AD$ by $M^{-1}(2)$.  A vector $N$ at $A$ becomes $M^{-1}(1) N$
at $D$, following the identification $AB \to DC$, and $M(2) N$
at $B$ following $BC \to AD$.  From $D$, the identification $BC$,
$AD$ leads to $M(2) M^{-1}(1) N$ at $C$, and from $B$ the
identification $AB \to DC$ leads to $M^{-1}(1) M(2) N$ at $C$.
The consistency condition is that the two ways of carrying $N$
to $C$ give the same result, in this case that $M(2) M^{-1}(1) N
= M^{-1}(1) M(2) N$.  This condition is equivalent to
$M(1) M^{-1}(2) M^{-1}(1) M(2) = I$, which is precisely the cycle
condition for genus one.

	Given the polygon, $p(0)$, and a timelike line at each corner, one
constructs the spacetime as follows.  One first constructs a family
of polygons $p(t)$ by sliding each corner of the polygon along the
corresponding timelike line, for a proper time $t$. For small enough
$t_0$, the region of Minkowski space which lies inside these
polygons for $0 \leq t \leq t_0$, is a truncated cylinder with
polygonal base; let its timelike walls be identified in pairs,
for instance the wall $E(1;t)$ is identified with the wall
$M^{-1}(1) E(1;t)$.  In this way one constructs a three-manifold
with topology $\Sigma_{g,N} \times (0,1)$ and no curvature, i.e.
a solution of $(2+1)$-dimensional gravity from $t = 0$ to $t = t_0$.
It can be shown that all solutions of $(2+1)$-dimensional gravity,
with the topology and sources considered here, can be obtained in
this way, and that the observable properties of the spacetime thus
constructed are independent of the choice of $N$ [21].

	In the example of genus one, the spacetime is defined by the
worldlines of the four corners, $A(t)$, $B(t)$, $C(t)$, $D(t)$,
$0 \leq t \leq t_0$.  Since the edges and diagonals are spacelike, by
hypothesis, there exists a spacelike surface which includes all
four points at a given t; it can further be shown that one can
choose a surface $\sigma_1(t)$ that is differentiable at the
identified edges [21].  Note that the choice of surface
$\sigma_1(t)$ is not an observable property of this spacetime.
On the other hand, the three-vector $E(1;t)$ $is$  observable:
If an observer at $A$ were to extend a straight stick of length
$\Vert E(1;t) \Vert $, in the direction of $E(1;t)$, the two ends
of the stick would touch.  To measure $M(1)$, the observer can
send a friend with a $(2+1)$-dimensional ``gyroscope" around the
same loop and measure its ``rotation" when he returns from
travelling around the loop. We emphasize that the polygon variables
are not associated to any particular choice of slicing of the
spacetime.

	If one considers one handle of a genus $g$ solution, the velocities
can be written in closed form as follows.  Let
$\mu = 1, 2$ be the two loops corresponding to this handle.  From
the brackets (3.1-3.2) and the constraints (3.4) one finds,
using the $SO(2,1)$-invariance of the structure constants
$\epsilon^{a b c}$,

$${dE(1)\over dt} \approx {1\over 2} \biggl( M(1) M(2)
M^{-1}(1) - I\biggr) N \eqno(3.10)$$

$${dE(2)\over dt} \approx {1\over 2} \biggl( M(2) M^{-1}(1) - M(1) M(2)
M^{-1}(1)\biggr) N \eqno(3.11)$$

This result can also be derived directly from the geometrical
picture which we have introduced above.  The right hand side of
equation (3.10) is the difference between the timelike image of $N$
at the tip of the segment $E(1)$ and the vector $N$, which we
have chosen to be based at the origin of $E(1)$.  Thus, the
constrained Hamiltonian system generates precisely the dynamical
evolution which we have given by constructing the spacetime from
a polygon and identifications.

	The picture which we have just presented, with the phase space
variables $E(\mu ;t)$ and $M(\mu)$, looks like an ordinary
classical system in parametrized form; in fact if it were not for
the constraints (3.4) and the mapping class group symmetry
(related to the arbitrary choice of $2g+N$ generators of the
homotopy group, [23]) , it would be closely related to a system
$2g + N$ free particles.  We will discuss next the realtion between
this formulations and the $ISO(2,1)$ homotopy invariants.  We will
first give the algebraic relation between the homotopies and the $E-M$
variables, then in sections 5 and 6 we will find the time-dependent
canonical transformation which relates Witten's invariant formalism to
the polygon representation.

\vfill
\eject

\noindent
\section{4. ISO(2,1) Homotopies from the Polygon Variables}

	A spacetime with topology $\Sigma_g \times (0,1)$ can be
represented by the $ISO(2,1)$ homotopies for $2g$ basis loops
$u_i, v_i$, as explained in Sec. 2.  The same spacetime can
be represented by a polygon embedded in $(2+1)$-dimensional
Minkowski space, as we saw in Sec. 3.  One handle of a slice
of the spacetime is represented by four edges of the polygon.
For the first handle, the edges are [Figure 4.1]: $E(1)$, $E(2)$,
$M^{-1}(1) E(1)$, and  $M^{-1}(2) E(2)$.  In this representation,
the loops $u_1$ and $v_1$ are represented as follows.  The loop $u_1$
is the path from an arbitrary base point $O$, to a point
$X$ on the edge $E(1)$, and continuing through the identified edge
$ - M^{-1}(1) E(1)$ to close the loop at $O$.  Similarly, $v_1$ is
a path from $O$ to $Y$ on the edge $- M^{-1}(2) E(2)$, continuing
from the identified point $Y'$ on the edge $E(2)$ to $O$ (the
reversed orientation for $v_1$ is conventional).

Having defined the loops $u_1$, $v_1$, it is now a straightforward
task to construct a representation of $\pi_1(\Sigma_g) \to ISO(2,1)$.
We begin with the homotopy $\rho (u_1)$.  This is a Poincar\' e
transformation, which can be represented as a four-by-four matrix
as (Sec. 2)

$$\rho(u_1) = \pmatrix{
M_1 & E_1\cr
0 & 1\cr}\eqno(4.1)$$

	The Lorentz projection of this homotopy, $M_1$, is the matrix
for parallel-transport along the loop back to the base-point $O$;
for the loop $u_1$ this is the matrix $M(1)$.  The translation
component $E_1$ is the path integral of the dreibein, parallel-
transported back along the path to $O$.  This is also the
displacement of an observer which travels once around the loop,
expressed in the frame at the base-point.  In the case of the
loop $u_1$, this is the displacement $OX + M(1) X'O$, where $X$
is the point at which the loop crosses the edge $E(1)$ and $X'$
is the identified point on the edge $-M^{-1}(1) E(1)$.  If we
denote by A the origin of the vector E(1), this deplacement can
be written as

$$\eqalign{
E & = OA + AX + M(1) (X^\prime D + DO)\cr
& = \biggl( I - M(1)\biggr) OA+M(1) \biggr(
M^{-1}(1)E(1)-E(1)-E(2)\biggr)\cr} \eqno(4.2)$$

\noindent
Thus, the $ISO(2,1)$ element for the loop $u_1$ based at O, is

$$\rho(u_1) = \pmatrix{
M(1) & (I-M(1))E(1)-M(1)E(2)+(I-M(1))OA\cr \ \ \cr
0 & \qquad \quad 1 \cr}\eqno(4.3)$$

Similarly, one finds

$$\rho(v_1) = \pmatrix{
M^{-1}(2) & (I-M^{-1}(1)-M^{-1}(2))E(1)\cr
& \qquad \quad  +(I-M^{-1}(2)) E(2) + (I-M^{-1}(2)) OA\cr \ \ \cr
0 & \qquad \qquad \qquad 1 \cr} \ \eqno(4.4)$$

	Each such $ISO(2,1)$ matrix leads to two Poincar\' e invariants:
the trace, $M^a_{\ a}$, and the scalar product of the translation
component with the ``rotation vector", $P^a = (1/2) \epsilon^{a b c}
M_{cb}$.  For a genus $g$ universe, there are $2g$ loops, $u_i$ and
$v_i$.  Taking all possible products of the corresponding
$ISO(2,1)$ matrices and constructing the corresponding invariants,
one obtains an infinite set of invariants.  Clearly they are not all
independent: for instance, $tr(M)$ and $tr(M^2)$ are related by
a Cayley-Hamilton-like identity [4].  In general, one can find
the complete set of relations among the different invariants, and
attempt to solve these constraints explicitly in order to identify
the reduced phase space.  In practice, this task is very difficult
[4], [5], [6], and has yet to be completed explicitly in the
general case (the number of independent invariants is $12g-12$,
the dimension of the reduced phase space).  We will return to
this problem in the case of genus two, where the reduced phase
space is  known explicitly [5].

	For the loop $u_1 v_1$, the two Poincar\' e invariants are

$$L_1 (u_1 v_1) = tr (M(2) M^{-1}(1)) - 1 \eqno(4.5)$$

$$L_2 (u_1 v_1) = E(1) \cdot P (2 {\buildrel \_ \over {1}}) +
E(2) \cdot P (2{\buildrel \_ \over {1}}) \eqno(4.6)$$

\noindent where

$$P^a (2{\buildrel \_ \over {1}}) = {1 \over 2} \epsilon^{abc}
(M(2)M^{-1}(1))_{cb} \eqno(4.7)$$

One may check that this is a constant of the motion by using
the expressions (3.10), (3.11) for the velocities:

$$\eqalign{ {dL_2(u_1 v_1) \over dt} & = P(2{\buildrel \_
\over{1}}) \cdot {d(E(1) + E(2)) \over dt} \cr
& = P(2{\buildrel \_ \over{1}}) \cdot
{ (M(2)M^{-1}(1) - I) N \over 2} = 0 \cr}
\eqno(4.8)$$

\noindent
since $P(2 {\buildrel \_ \over{1}} )$ is invariant under
$M(2) M^{-1}(1)$.

	The $ISO(2,1)$ homotopy for a loop which surrounds a particle
is computed as follows.  Let $w_1$ be the loop which goes from $O$,
to a point $X$ on the edge $E(2g+1)$, then back to $O$ through the
corresponding point $X'$ on the identified edge $M^{-1}(2g+1) E(2g+1)$.
Proceeding as above, one obtains the homotopy

$$\rho(w_1) = \pmatrix{
M(2g+1) & (I-M(2g+1)) E(2g+1) \cr
\ & \qquad + (I-M(2g+1))OF\cr \ & \ \cr
0 & \qquad \qquad 1 \cr}\eqno(4.9)$$

\noindent where $F$ is the base point of the three-vector $E(2g+1)$.
Note that since $A$ was defined as the base point of $ E(1)$, and
the vectors $E(1)$, $E(2)$, $- M^{-1}(1) E(1)$, ..., $- M^{-1}(2g)
E(2g)$ represent the section of polygon which lies between $A$ and
$F$, one has

$$OF = OA + AF \eqno(4.10)$$

$$AF = (I - M^{-1}(1) E(1) + \cdots + (I-M^{-1}(2g))
E(2g)\eqno(4.11)$$

$$ \ \ = J(1) + J(2) + \cdots + J(2g)\eqno(4.12)$$

\noindent where $J(\mu) = E(\mu) - M^{-1}(\mu) E(\mu)$.

	From the geometric picture in which we have represented both
the $ISO(2,1)$ homotopies and the ($E-M$) variables, one expects
that the cycle condition for the $ISO(2,1)$ homotopies,

$$\eqalign{ (\rho(u_1) & \rho(v_1) \rho (u_1^{-1})
\rho(v_1^{-1})) \cdots (\rho(u_g) \rho(v_g) \cr
& \rho(u_g^{-1}) \rho(v_g^{-1})) \rho(w_1) \rho(w_2) \cdots
\rho(w_N) \approx I\cr} \eqno(4.13)$$

\noindent would be equivalent to the six Poincar\' e constraints,

$$J \equiv \sum_\mu (I-M^{-1}(\mu)) E(\mu) \approx 0 \eqno(4.14)$$

$$P^a \equiv {1\over 2} \epsilon^{abc} W_{cb} \approx 0 \eqno(4.15)$$

	To check this explicitly, one simply computes the product of
$ISO(2,1)$ matrices in the order (4.13), using the relations (4.3),
(4.4), etc.  This takes a bit of work, but the result is simple:
one finds

$$\rho(u_1) \rho(v_1) \rho(u_1^{-1}) \rho(v_1^{-1}) \cdots
 \rho(w_N) = \left( \matrix{
W & -WJ + (I-W)OA\cr \ & \ \cr
0 & 1 \cr} \right) \eqno(4.16)$$

\noindent where W is the $SO(2,1)$ ``cycle":

$$\eqalign{& W = \biggl( M(1) M^{-1}(2) M^{-1}(1) M(2)\biggr)
\cdots \biggl( M(2g-1) M^{-1}(2g) \cr
& M^{-1}(2g-1) M(2g)\biggr) M(2g+1) M(2g+2) \cdots
M(2g+N)\cr} \eqno(3.5)$$

	In the next sections, we will choose a
definition of internal time, compute the corresponding Hamiltonian,
then carry out the Hamilton-Jacobi transformation explicitly.

\vfill
\eject

\section{5. Time}

 Since the variables $E(\mu ; \tau )$ depend linearly on the
evolution parameter $\tau $, a natural choice for an internal
time variable, or ``clock", is any one component of such a vector,
such as $E^x(1)$.  We will consider separately the closed universes
with $N = 0$, $g \geq 2$, then the closed and open universes with
$N \geq 2$, $g \geq 0$.  The other cases can be treated in a similar
fashion, except for the torus ($N = 0, g = 1$) which will be discussed
in Sec. 7.1.

\noindent 5.1 Time and the Hamiltonian for Empty Genus $g$ Universes

	For $N = 0$, the $SO(2,1)$ projections of the $ISO(2,1)$
homotopies generate a Fuchsian group [24]; this implies that
the $SO(2,1)$ matrices {$M(1), \cdots $} are boosts.  Therefore, one
can choose a frame where $P(1)$
is parallel to the $x$-axis.

$$P^t(1) = 0\eqno(5.1)$$

$$P^y(1) = 0 \eqno(5.2)$$
\noindent and solve these simultaneously with the corresponding
constraints,

$$J^t = 0 \eqno(5.3)$$

$$J^y = 0 \eqno(5.4)$$

\noindent In this gauge, $M(1)$ is a pure boost in the $y-t$-plane,

$$M(1) = \pmatrix{cosh (b) & 0 & sinh (b)\cr
0 & 1 & 0 \cr
sinh (b) & 0 & cosh (b)\cr} \eqno(5.5)$$

\noindent	and the constraints $J^t = J^y = 0$,

$$J \equiv \biggl( I-M^{-1}(1)\biggr) E(1) + \sum_{\mu>1} \biggl( I-M^{-1}
(\mu)\biggr) E(\mu) \approx 0 \eqno(5.6)$$

\noindent split into the $x$-component, for which the first term
vanishes thanks to (5.5),

$$J^x \equiv \sum_{\mu > 1} \biggl( I-M^{-1}(\mu)\biggr)^x_{\ a} E^a(\mu)
\approx 0 \eqno(5.7)$$

\noindent and two equations which we solve for $E^t(1)$ and $E^y(1)$:

$$\pmatrix{E^t (1)\cr \ \cr
E^y (1) \cr} =  \pmatrix{ -{1\over 2} & {sinh (b) \over 2(1-cosh
(b))} \cr & \ \ \cr
{sinh (b) \over 2(1-cosh (b))} & - {1\over 2} \cr}
\pmatrix{ J_1{}^t\cr & \ \ \cr
J_1{}^y\cr} \eqno(5.8)$$

\noindent where

$$J_1 \equiv \sum_{\mu > 1} (I-M^{-1}(\mu)) E(\mu) \eqno(5.9)$$

\noindent	With this choice of frame, the component $E^x(1)$, which
we would like to use as a clock, is canonically conjugate to the
boost parameter $b$.  Indeed,

$$\{ E^a(\mu), P^b(\mu) \} = {1\over 2} \biggl( \eta^{ab} tr M(\mu) -
M^{ba}(\mu) \biggr) \eqno(5.10)$$

\noindent so that, for $a = b = x$ and $\mu = 1$, one finds

$$\{ E^x (1), P^x(1) \} = cosh (b) \eqno(5.11)$$

\noindent or, using $P^x(1) = sinh(b)$,

$$\{ E^x(1), b \} = 1 \eqno(5.12)$$

\noindent $P^x(1)$ is not bounded from below, but if we choose
as internal time

$$T = - {E^x (1) \over P^x (1)} \eqno(5.13)$$

\noindent then the conjugate variable is $P_T = -cosh(b)$ and
the Hamiltonian is

$$H = - P_T = {1\over 2} \biggl( tr M(1) - 1 \biggr) \eqno(5.14)$$

\noindent
which is  bounded from below.  This is of course one of many possible
choices for the time/hamiltonian pair. In $(2+1)$-dimensional
gravity there is no outstanding choice of ``clock". In the classical
theory all choices of time lead to the same equations of motion,
but this is not generally true in the quantum theory [25], [26];
we will discuss this point further in the second article.

	The Hamiltonian (5.14) is given in terms of the variables $M(\mu)$,
$\mu > 1$, by solving the constraint

$$W = \biggl( M(1) M^{-1}(2)M^{-1}(1)M(2)\biggr) \cdots \biggl(
M(2g-1) M^{-1}(2g)M^{-1}(2g-1)M(2g)\biggr) \eqno(5.15)$$

\noindent for $M(1)$ (a boost of magnitude $b$, with axis $x$).
One finds

$$H = {P_t(2) Q_t - P_y(2) Q_y \over P_y^2(2) - P_t^2(2)} \eqno(5.16)$$

\noindent where

$$\eqalign{ Q^a & = {1 \over 2} \epsilon^{abc} \biggl( M(2) (M(3)
M^{-1}(4)M^{-1}(3)M(4)) \cdots \cr
& (M(2g-1) M^{-1}(2g) M^{-1}(2g-1) M(2g))\biggr)_{cb}\cr}\eqno(5.17)$$

Altogether, we have imposed three gauge conditions out of six
Poincar\' e symmetries.  The remaining phase space variables are
{$E(\mu), M(\mu)$;\  $\mu = 2, ..., 2g$}.  Their time-evolution is
generated by the Hamiltonian (5.16), and they are subject to the
three remaining first class constraints, which commute with the
Hamiltonian. These are

$$J^x \equiv \sum_{\mu > 1} \biggl( I-M^{-1}(\mu) \biggr)^x_{\ a}  \ E^a(\mu)
\approx 0 \eqno(5.18)$$

and, from $M(1) M^{-1}(2) M^{-1}(1) \approx \biggl(M(2) M(3)
M^{-1}(4) M^{-1}(3) M(4) ... M(2g) \biggr) ^{-1}$,

$$P^x (2) - Q^x \approx 0 \eqno(5.19)$$

$$P^2 (2) - Q^2 \approx 0 \eqno(5.20)$$

	One could fix the remaining three gauge symmetries, which are
generated by (5.19)-(5.20), and eliminate $E(2)$ and $M(2)$,
thereby reducing the theory to an ordinary Hamiltonian system
with $6 \times (2g-2)$ independent phase space variables.
However, this would be of little practical interest in what
follows, and would make the formalism significantly more cumbersome.

	Using the Hamiltonian (5.16), one can compute the velocities of
any function on phase space.  For example, consider the edge vector
$E(3)$ of the polygon:  By computing the bracket with the
Hamiltonian and then using the constraints (5.15), one finds

$${dE(3)\over dt} = {(M^{-1}(2)-M(3)M(4)M^{-1}(3)M^{-1}(2)) (TrM(2) -
M(1)M^{-1}(2)M^{-1}(1))\over 2(P_y^2(2) - P_t^2(2))} {\cal P}_x P(2)
\eqno(5.21)$$

\noindent where

$${\cal P}_x = \pmatrix{
1 & 0 & 0 \cr
0 & 0 & 0 \cr
0 & 0 & 1 \cr} \eqno(5.22)$$

	This velocity can also be computed in the parametrized formalism
defined in Sec. 3; this allows us to compute the lapse-shift
vector which corresponds to our choice of gauge: one finds that
$dE(3)/dt = N^a \{ P_a, E(3) \}$ when $N$ is given by

$$N = \biggl({M(1)M^{-1}(2)M^{-1}(1) - TrM(2) \over P^2_y(2) -
P_t^2(2)} \biggr)
M(1)M^{-1}(2)M^{-1}(1) {\cal P}_x P(2)\eqno(5.23)$$

$$P_a \equiv {1\over 2} \epsilon^{abc} W_{cb} \approx 0 \eqno(5.24)$$

$$W = \biggl( M(1) M^{-1}(2)M^{-1}(1)M(2)\biggr) \cdots \biggl(
M(2g-1) M^{-1}(2g)M^{-1}(2g-1)M(2g)\biggr) \eqno(5.25)$$

	Of course if one had chosen any other $E(\mu)$ to carry out
this calculation, one would have found the same lapse-shift
vector $N$.  For instance, one can show (with some work)
that the velocity $dE(2)/dt = N^a \{ P_a, E(2) \}$ is equal
to $\{ H, E(2) \}$ when $N$ is given by (5.23).

\noindent 5.2 Closed Universes with $N \geq 2$ Particles

	Since the methods and derivations for open and closed universes
with particles are conceptually identical to those of the first
sub-section, we will run through the next cases rapidly and refer
the reader to the literature for details.  The motion of any
particle in an otherwise unspecified geometry can be obtained
from the Hamiltonian (2.7). One finds, for $\mu = 2g+1$ and
$N(2g+1) = N^t/(4cos \Omega (2g+1) P^t(2g+1))$  (see ref. [12]),

$${dE(2g+1)\over dt} = \biggl( 0, - {P^x(2g+1)\over P^t(2g+1)} , -
{P^y(2g+1)\over P^t(2g+1)} \biggr) \eqno(5.26)$$

Similarly for particle $\mu =2g+2$, one may choose a
parametrization of its worldline,  $N(2g+2)$, that is such that
its velocity lie in the $(x-y)$-plane.  One chooses a frame where
$M(2g+2)$ is a pure rotation ($P^x(2g+2) = P^y(2g+2) = 0$),
and uses the residual freedom to rotate around the $t$-axis to
set $P^y(2g+1) = 0$.  Combining these gauge choices, the observer
is at rest with respect to particle $(2g+2)$, and sees particle
$(2g+1)$ moving along the $x$-axis at a constant velocity.  The
$x$-component of the second particle could then be chosen as ``clock",
however this would complicate the task of obtaining the Hamiltonian
function, because these gauge choices require computing Dirac
brackets between the matrix $M(2g+1)$ and the vector $E(2g+1)$.
If we choose, instead, to use the $x$-component of the canonical
variable $X(2g+1)$ (Sec. 6.1) and $T = X^x(2g+1) / P^x(2g+1)$,
one finds, as before, $H = tr(M(2g+1))$.  $H$ must be calculated as a
function of the other matrices, by solving the cycle conditions for
$M(2g+1)$, given that $M(2g+2)$ is a pure rotation with angle
$\Omega (2g+2)$.

$$H=Tr \biggl( M(2g+2) \cdots M(2g+N) M(1)M^{-1}(2)M^{-1}M(2)\cdots
M^{-1}(2g-1) M(2g)\biggr) \eqno(5.27)$$

\noindent where

$$M(2g+2) = \pmatrix{ 1 & 0 & 0 \cr \ &\ & \ \cr
0 & cos(\Omega(2g+2)) & sin(\Omega(2g+2))\cr \ &\ & \ \cr
0 & -sin(\Omega(2g+2)) & cos(\Omega(2g+2))\cr} \eqno(5.28)$$

\vfill
\eject

\noindent 5.3 Open Universes with $N \geq 2$ Particles

	If the universe has the topology $\Real^2$, then the geometry at
infinity approximates that of a cone with a helical shift [27].  In
this case, there is a well-defined direction for the flow of time,
which is given by the axis of the cone.  The deficit angle of
this cone is the total energy of the universe, and the helical
shift its total angular momentum.  The $SO(2,1)$ ``cycle", which
was required to be equal to the identity matrix in order that the
universe may close, is now a matrix which is generally $not$ equal to
the identity, but corresponds to parallel transport around a
loop at infinity.  It is usually required that its axis be timelike,
in which case there are no closed timelike curves [28], [29]; it
is then a rotation and the angle of this rotation is the deficit
angle which corresponds to the conical geometry at infinity, i.e.,
the energy.  So the Hamiltonian is the following function of the
matrices $M(\mu)$:

$$H = Ar cos \biggl( {trW-1\over 2} \biggr) \eqno(5.29)$$

\noindent where W is the cycle, equal to the left-hand side of
Equation (3.7) (but it is no longer required to be trivial: $W \not= I$).

\vfill
\eject

\section{6. The Hamilton-Jacobi Transformation}

	We will only consider the case 5.1 ($N = 0, g \geq 2$); the
generalization to $N \not= 0$ and $g \geq 0$ is straightforward.
We will perform three successive changes of variables, to relate
the time-dependent variables {$E(\mu;T), M(\mu), \mu = 2,..., 2g$}
to the homotopy invariants.  The first transformation is a change
to canonical variables {$X(\mu;T), P(\mu)$}.  We then perform
the Hamilton-Jacobi transformation, with the internal time which
we introduced in the previous section.  Finally, the homotopy
invariants will be computed from the Hamilton-Jacobi variables.

\noindent 6.1 Canonical Variables

	The canonical phase space variables are defined as follows.

$$X(\mu) = {1\over P^2(\mu)} \biggl( P(\mu) \wedge J(\mu) - {2(E(\mu)
\cdot P(\mu)) P(\mu) \over tr M(\mu)-1} \biggr) \eqno(6.1)$$

\noindent where, by definition,

$$J(\mu) = (I-M^{-1}(\mu)) E(\mu)\eqno(6.2)$$

$$\bigg( A \wedge B\biggr)^a = \epsilon^{abc} A_b B_c\eqno(6.3)$$

\noindent and the ``momentum" variables are

$$P_a(\mu) = {1\over 2} \epsilon_a{}^{bc} M_{cb}(\mu) \eqno(6.4)$$

\noindent One may check that the brackets of the variables
$X(\mu), P(\mu)$ are canonical, by using (3.1)-(3.2):

$$\{ P^a(\mu), \ X^b(\mu) \} = \delta_a^b \eqno(6.5)$$

	The change of variables defined by (6.1)-(6.4) is regular.
The inverse is given by either one of the following two sets of
relations, depending on $M(\mu)$.  If $M(\mu)$ is a boost, then

$$E(\mu) = {M^{1/2} P \wedge (P\wedge X) \over P
\sqrt{2\sqrt{P^2+1}-2}} - {\sqrt{P^2+1}(P \cdot X)P\over P^2} \eqno(6.6)$$

\noindent where the index $\mu = 1,2,...,2g$ was omitted in
the right hand side, for clarity, and

$$\eqalign{ (M^{1/2})^a{}_b = & \delta^a{}_b +
{ \sqrt{{ \sqrt{P^2+1}-1} \over 2P^2}} P_c \epsilon^{ca}{}{}_b\cr
& +\biggl( \sqrt {{ \sqrt{P^2+1}+1}\over 2} - 1 \biggr)
\biggl(\delta^a{}_b - {P^aP_b\over P^2}\biggr)\cr} \eqno(6.7)$$

\noindent If $M(\mu)$ is a rotation, as for particles,

$$E(\mu) = {M^{1/2} P\wedge (P\wedge X) \over P \sqrt{2 -2
\sqrt{1-P^2}}} - {\sqrt{1-P^2}(P\cdot X)P\over P^2} \eqno(6.8)$$

$$\eqalign{ (M^{1/2})^a{}_b & = \delta^a{}_b +
\sqrt{{ {1-\sqrt{1-P^2}} \over 2P^2 }} P_c \epsilon^{ca}{}{}_b \cr
& + \bigg( \sqrt {{ \sqrt{1-P^2} +1} \over 2} -1 \biggr) \biggl(
\delta^a{}_b - {P^a P_b \over P^2} \biggr) \cr}\eqno(6.9)$$

The matrices $M(\mu)$ are given as a function of $P(\mu)$ as
follows.  For a boost,

$$M^a{}_b = \delta^a{}_b + P_c \epsilon^{ca}{}{}_b + \biggl(
\sqrt{1+P^2}-1 \biggr) \biggl( \delta^a{}_b - {P^a P_b\over P^2}
\biggr) \eqno(6.10)$$

\noindent and for a rotation,

$$M^a{}_b = \delta^a{}_b + P_c \epsilon^{ca}{}{}_b + \biggl(
\sqrt{1-P^2} -1 \biggr) \biggl(\delta^a{}_b - {P^a P_b\over P^2}
\biggr) \eqno(6.11)$$

      The constraints $J \approx 0,  P \approx 0$ can be written in terms
of the canonical variables, using (6.6)-(6.11) and

$$(I-M^{-1}(\mu)) E(\mu) = J(\mu) = X(\mu) \wedge P(\mu)
\eqno(6.12)$$

\noindent The $SO(2,1)$ constraints become simply

$$\sum_\mu  J(\mu) \approx 0 \eqno(6.13)$$

\noindent while the translation constraints $P \approx 0$ are
defined implicitly in terms of {$P(\mu)$} by (6.10) and (6.11).
The explicit form of the function $P(P(\mu )) \approx 0$ can be
computed explicitly.

\noindent 6.2 The Hamilton-Jacobi Transformation

	Having found the Hamiltonian in the previous section, it is a
straightforward task to carry out the Hamilton-Jacobi transformation
from the $X(\mu ), P(\mu )$ variables.  The generating function
is given in terms of the new momenta $P'(\mu )$ and the old
variables $X(\mu )$ as the ``identity operator" minus the ``time
evolution operator":

$$G(X, P^\prime; T) = P^\prime \cdot X - \int^T_0 \ H(P) dt \eqno(6.14)$$

\noindent where the old momenta in the Hamiltonian function must
be replaced in terms of the new ones by inverting the relations
which define the canonical transformation, namely

$$P(\mu) = {\partial G\over \partial X(\mu)} \eqno(6.15)$$

$$X^\prime (\mu) = {\partial G \over \partial P^\prime(\mu)}\eqno(6.16)$$

	Since the Hamiltonian is a function of P's only, the task of
replacing $P(\mu )$ in terms of $P'(\mu )$, usually the main
difficulty in deriving the generating function, is trivial:
$P(\mu ) = P'(\mu )$.  The integral over time is also easy,
since the Hamiltonian does not depend explicitly on time.  Thus,

$$G (X, P^\prime; T) = P^\prime \cdot X - \biggl(
{P^\prime_t(2)
Q^\prime_t - P^\prime_y (2) Q^\prime_y \over P^{\prime 2}_y (2) -
P^{\prime 2}_t (2)} \biggr) T \eqno(6.17)$$

\noindent where

$$Q^{\prime a} = {1\over 2} \epsilon^{abc} \biggl(
M^\prime(2)M^\prime(3) M^{\prime -1}(4) M^{\prime -1}(3) M^\prime(4)
\cdots M^\prime(2g)\biggr)_{cb} \eqno(6.18)$$

	The new Hamiltonian is $K = H + \partial{G} / \partial{T} = 0$,
and the new canonical variables $X'(\mu ), P'(\mu )$ are given
in terms of the old ones by the equations (6.15)-(6.16), i.e.
$P'(\mu) = P(\mu)$ and

$$X^\prime (\mu) = X(\mu) - V(\mu) T \eqno(6.19)$$

\noindent where

$$V(\mu) = \{ {P_t(2) Q_t - P_y(2) Q_y \over P_y^2(2) -
P_t{}^2(2)}, X(\mu) \} \eqno(6.20)$$

In these expressions, one should remember that $Q$ and $P(2)$
are related by the first-class constraints (5.25).

\noindent 6.3 Time Evolution in the Hamilton-Jacobi Variables

	The Hamilton-Jacobi variables $X'(\mu ), \mu >1$, are constants
of the motion.  In fact, one goes from the old variables to the
new ones precisely by subtracting the time evolution out of
$X(\mu ;T)$, so that $X'(\mu ) = X(\mu ;0)$.  In order to derive
the time-evolution of the dynamical variables, one inverts the
relations which define the Hamilton-Jacobi variables in terms
of the time-dependent ones:

$$X(\mu ; T) = X^\prime(\mu) + \{ H, X(\mu) \} T \eqno(6.21)$$

Of course this is the usual way that the time evolution is
extracted from the Hamilton-Jacobi formalism, only that it
appears particularly simple in this case because the Hamiltonian
depends only on the momenta.  We will show next that the homotopy
invariants are functions of the Hamiton-Jacobi variables which
we have just derived.

\

\

\noindent 6.4 Homotopy Invariants as Functions of the
Hamilton-Jacobi Variables

	As we have seen, the homotopy invariants can be separated in two
types (Equations 4.5-4.6).  The first are traces of the Lorentz
projections of the $ISO(2,1)$ homotopies, such as:

$$L_1 (u_1 v_1) = tr \biggl( M(2) M^{-1}(1) \biggr) - 1 \eqno(6.22)$$

The other are the scalar products of the vector components of the
$ISO(2,1)$ homotopies, with the duals of their Lorentz projections,
such as:

$$L_2 (u_1 v_1) = E(1) \cdot P(2{\buildrel \_ \over {1}}) + E(2)
\cdot P(2 {\buildrel \_ \over {1}}) \eqno(6.23)$$

	The Poisson brackets of these observables can be computed
directly using the brackets of the polygon variables $E(\mu ), M(\mu )$.
For instance, one finds with some algebraic work

$$\{ L_1 (\rho), L_1(\rho^\prime) \} = 0 \  \forall\rho , \rho^\prime \in
\pi_1 (\Sigma_g) \eqno(6.24)$$

$$\{ L_1(u_1), L_2(v_1) \} = L_1 (u_1 v_1) - L_1 (u_1 v_1^{-1}) \eqno(6.25)$$

$$\{ L_2 (u_1), L_2(v_1) \} = L_2 (u_1 v_1) - L_2 (u_1 v_1^{-1}) \eqno(6.26)$$

These are the usual expressions for the brackets of the homotopy
invariants, when they are calculated directly from the field theory
(see, e. g., Nelson and Regge [17]).

	In the expressions (6.22)-(6.23), one identifies the explicit
dependence on time by replacing $E(1)$ by its expression (5.13)
from gauge-fixing. For example,

$$L_2 (u_1 v_1) = E(2) \cdot P(2 {\buildrel \_ \over {1}}) + A P^t(2
{\buildrel \_ \over {1}}) - B P^y(2 {\buildrel \_ \over {1}})
- P^x(1) P^x(2 {\buildrel \_ \over{1}}) T \eqno(6.27) $$

$$\eqalign{ = E^\prime (2) \cdot P^\prime (2 {\buildrel \_ \over
{1}}) + A^\prime P^{\prime t} (2{\buildrel \_ \over{1}}) - B^\prime
P^{\prime y} (2 {\buildrel \_ \over{1}})} \eqno(6.28)$$

\noindent where

$$A = {(1-cosh(b)) J_1{}^t + sinh(b) J_1^y \over 2(1- cosh(b))} ,
\eqno(6.29)$$

$$B = {sinh (b) J_1{}^t  + (1 - cosh (b)) J_1{}^y \over 2 (1-
cosh(b))}, \eqno(6.30)$$

$$J_1 = \sum_{\mu > 1} \biggl(I - M^{-1}(\mu)\biggr) E(\mu)
\eqno(6.31)$$

Equation (6.28) gives $L_2(u_1 v_1)$ in terms of the Hamilton-
Jacobi variables {$X'(\mu ), P'(\mu ), \mu =2,..., 2g+N$}, while
(6.27) gives the same invariant in terms of the dynamical
variables and time.  All of the homotopy invariants can be similarly
expressed in terms of the Hamilton-Jacobi variables.  Since these
form a complete description of the phase space, the expressions
can be inverted to give the Hamilton-Jacobi variables as functions
of the homotopy invariants.  The time-evolution problem is solved
by expressing the invariants as explicit functions of time.

	Altogether, we have gone from the $E(\mu ), M(\mu )$ variables
to $X(\mu ), P(\mu )$ with a  change of variables, then applied
the Hamilton-Jacobi transformation, and finally with another
regular change of variables we recovered the homotopy invariants.
The {$E(\mu ), M(\mu )$} variables, modulo the  constraints
and gauge symmetry, span the same reduced phase space as the homotopy
invariants, which shows that the two formulations of $(2+1)$-
dimensional gravity are closely related, at least classically.  We
now turn to two examples.  The example of the torus is reasonably
well understood already [7]; we use it mainly to throw a bridge
with existing literature.  For the genus two example,
on the other hand, this is the first explicit solution of the
Hamilton-Jacobi problem.

\vfill
\eject

\section{7. Examples}

	Most recent work on $(2+1)$-dimensional gravity (classical and
quantum) has focused on spacetimes with the topology $T^2 \times
(0,1)$, largely because this is the case where the time evolution
problem has been solved explicitly in York's extrinsic time [7].

\noindent The reader interested in a clear illustration of the
arguments of the previous sections should turn to the second example
(genus 2), which is more representative of the general case.
This subsection is designed specifically for those who have worked,
or have the intention of working, on the problem of $T^2 \times
(0,1)$ gravity.

	We will review the example of the torus in our formalism,
emphasizing its particularities, then compare with the literature.

	The reduced variables for the torus are two three-vectors $E(1),
E(2)$, and two Lorentz matrices, with the constraints

$$J \equiv E(1) + E(2) - M^{-1}(1) E(1) - M^{-1}(2) E(2) \approx 0
\eqno(7.1)$$

$$P^a \equiv {1 \over 2} \epsilon^{abc} \biggl( M(1) M^{-1}(2) M^{-1}(1)
M(2) \biggr)_{cb} \approx 0 \eqno(7.2)$$

The first constraint states that the four vectors in the sum
form a closed quadrilatteral [Figure 7.1].  We choose two loops
which form a basis of the homotopy group, as follows.  One, $u$,
begins from a point $O$ inside the quadrilatteral, crosses through
the edge $E(1)$ at $X$ and returns to $O$ from the identified
point $X'$.  The other, $v$, crosses through the edge $M^{-1}(2)
E(2)$ at $Y$ and continues through the identified point $Y'$.
The corresponding $ISO(2,1)$ homotopies are

$$\rho(u) = OX + M(1) X^\prime 0 = \pmatrix{
M(1) & -M(1)M^{-1}(2)E(2) + (I-M(1))OA \cr \ &\ \cr
0 & 1 \cr} \eqno(7.3)$$

$$\rho(v) = \pmatrix{
M^{-1}(2) & -M^{-1}(2)E(1) + (I-M^{-1}(2))OA\cr \ &\ \cr
0 & 1 \cr} \eqno(7.4)$$

	These matrices generate a representation of $\pi_1(\Sigma_g)
\to ISO(2,1)$ and, consequently, they satisfy the cycle condition.

$$\rho(u) \rho(v) \rho(u^{-1}) \rho(v^{-1}) = I \eqno(7.5)$$

Indeed, computing the product (7.5) explicitly, one finds

$$\pmatrix{
M(1)M^{-1}(2)M^{-1}(1)M(2) & - (M(1)M^{-1}(2)M^{-1}(1)M(2)) J\cr
& + (I-M(1)M^{-1}(2)M^{-1}(1)M(2)) OA\cr \ &\ \cr
0 & 1\cr} \eqno(7.6)$$

This identity states that the closed path $OX + X'O + OY + Y'O -
OX' - XO - OY' - YO$ is contractible, i.e. belongs to the trivial
homotopy class [Figure 7.1].  Note that the cycle condition for
the $ISO(2,1)$ homotopies summarizes both the cycle condition for
$SO(2,1)$ homotopies and  the closure condition for the polygon.

The number of independent observables should be equal to the
dimension of the cotangent bundle of Teichmller space, namely 4.
For the torus, the $SO(2,1)$ cycle condition states that the
matrices $M(1)$ and $M(2)$ commute, only two independent
conditions.  Also, given that these matrices commute (and
therefore have parallel axes), the dot product of the closure
condition with either one of the rotation vectors, say $P(1)^.J$,
is identically zero, since $P(1)$ is parallel to the common axis
of $M(1)$ and $M(2)$, and

$$P(1) \cdot J = P(1) \cdot \biggl( I - M^{-1}(1) \biggr) E(1) + P(1)
\cdot \biggl( I - M^{-1}(2)\biggr) E(2) \equiv 0 \eqno(7.7)$$

So the number of independent constraints is really four,
not six, and the number of degrees of freedom is $12 - 2 \times 4
= 4$.

	The four homotopy invariants are

$$L_1 (u) = Tr M(1) \eqno(7.8)$$

$$L_2 (u) = P(1) \cdot E(2) \eqno(7.9)$$

$$L_1 (v) = Tr M(2) \eqno(7.10)$$

$$L_2 (v) = P(2) \cdot E(1) \eqno(7.11)$$

	We will fix the frame by setting

$$P^y(1) = P^t (1) = 0 \eqno(7.12)$$

\noindent and solving the two independent constraints $J \approx 0$ for

$$\pmatrix{E^t (1)\cr \ \cr
E^y (1) \cr} =  \pmatrix{ -{1\over 2} & {sinh (b) \over 2(1-cosh
(b))} \cr & \ \ \cr
{sinh (b) \over 2(1-cosh (b))} & - {1\over 2} \cr}
\pmatrix{ J^t(2)\cr & \ \ \cr
J^y(2)\cr} \eqno(7.13)$$

This leaves the variables $E^x(1)$ and $E^a(2)$, a=0,1,2, and the
constraints $P^a \approx 0$ which are now equivalent to

$$P^y(2) \approx 0 \eqno(7.15)$$

$$P^t(2) \approx 0 \eqno(7.16)$$

The internal time must be a variable which does not commute with
these constraints - the choice $T = -E^x(1) / P^x(1)$ analogous to
that of Sec. 4 does not satisfy this criterion.  One may choose
instead, for example, $T = -(cos(\alpha) X^x(2) + sin(\alpha)
X^y(2)) $, where $\alpha$ is a parameter not equal to zero.  This
fixes the gauge freedom for the constraint $P_y = 0$.  There
is one remaining gauge freedom, which one can fix
by choosing the gauge condition $X^t(2) = 0$.  The Hamiltonian
corresponding to $T$ is

$$H = cos \alpha P^x(2) + sin \alpha P^y(2)
\approx cos \alpha \ P^x(2) \eqno(7.17)$$

 For $\alpha = \pi / 2$, the Hamiltonian vanishes and the reduced
phase space variables $X^x(1)$, $X^x(2)$, $P^x(1)$ and $P^x(2)$
are Hamilton-Jacobi invariants.

	We close this section with a short dictionnary to translate
between our results and other authors' results for the torus (in the
extrinsic curvature time).

	The extrinsic time is alternatively written $\tau = tr(K)$, or $q_0$.

	The moduli are written as a complex number $m$, or specifically as
$m = m_1 + i m_2 = q_1 + i q_2$.  They are conjugate to $p = p_1 +
i p_2$.  The moduli and their momenta are given in terms of the
$ISO(2,1)$ homotopies $\rho (u) = \Lambda_1$ and $\rho (v) =
\Lambda_2$ by the relations [10]

$$m = \biggl( a+ {i \lambda\over \tau} \biggr)^{-1} \biggl( b +
{i\mu\over \tau}\biggr) \eqno(7.18)$$

$$p = - i\tau \biggl( a - {i\lambda \over \tau}\biggr)^2 \eqno(7.19)$$

\noindent where $a, b, \lambda, \mu$  are defined by

$$\rho(u) = \pmatrix{
cosh \lambda & 0 & sinh \lambda & 0 \cr
0 & 1 & 0 & a \cr
sinh \lambda & 0 & cosh \lambda & 0 \cr
0 & 0 & 0 & 1\cr} \eqno(7.20)$$

$$\rho(v) = \pmatrix{
cosh \mu & 0 & -sinh \mu & 0 \cr
0 & 1 & 0 & b \cr
-sinh \mu & 0 & cosh \mu & 0 \cr
0 & 0 & 0 & 1 \cr} \eqno(7.21)$$

This gives us the correspondence with our variables at $t = 0$.
Using (4.3)-(4.4), one finds

$$E(1{;} \ t = 0 ) = (0, -b, 0) \eqno(7.22)$$

$$M(1) = \pmatrix{
cosh \lambda & 0 & sinh \lambda \cr
0 & 1 & 0 \cr
sinh \lambda & 0 & cosh \lambda \cr}\eqno(7.23)$$

$$E(2{;} \ t = 0) = (0, -a, 0) \eqno(7.24)$$

$$M(2) = \pmatrix{
cosh \mu & 0 & -sinh \mu \cr
0 & 1 & 0 \cr
-sinh\mu & 0 & cosh\mu \cr} \eqno(7.25)$$

Note that the solution is singular at $t = 0$, in the sense
that the torus has vanishing area there.  One easily checks that
the area is positive for all $t \not= 0$, by constructing the
spacetime as explained in the beginning of this section.

	The extrinsic time $\tau$, itself, is a function of the polygon
variables:  It is the trace of the extrinsic curvature of a constant
curvature slice which must respect the identification conditions.
This is a straightforward but lengthy exercise in geometry. One finds

$$\tau= \sqrt{ {(E(1) \cdot E(2)) A+E^2(1) B+E^2(2) C+ \sqrt{\Delta}}
\over 4A^2 - 16 BC} \eqno(7.26)$$

\noindent where

$$\eqalign{A = cosh (\mu) & cosh (\mu+\lambda) - cosh (\mu) -
cosh (\mu +\lambda) - 1 \cr
& - sinh (\mu+\lambda) sinh (\mu) \cr} \eqno(7.27)$$

$$B = cosh(\mu) \  cosh (\mu + \lambda) - 1 - sinh(\mu) \  sinh(\mu+
\lambda) \eqno(7.28)$$

$$C = cosh(\mu) - 1 \eqno(7.29)$$

$$\eqalign{\Delta & = (A^2 - BC) E^2(1) E^2(2) + 4BC (E(1) \cdot E(2))^2\cr
& + B^2 E^4(1) + C^2 E^4(2) + 2 (E(1) \cdot E(2)) (ABE^2(1) +
ACE^2(2))\cr} \eqno(7.30)$$

	The Hamiltonian which corresponds to this choice of time is given,
in various notations, as

$$H = {\sqrt{q_2^2 (p_1^2 + p_2^2)} \over q_0} \eqno(7.31)$$

$$H = {\vert a\mu - b\lambda \vert \over \tau} \eqno(7.32)$$

$$ H = {(E(1) \cdot P(1)) sinh^{-1} (P(1))\over \tau \ P(1)} + {(E(2)
\cdot P(2)) sinh^{-1} (P(2)) \over \tau \ P(2)} \eqno(7.33)$$

	Finally, the invariants are

$$E(1) \cdot P(2) = - a \ sinh \lambda \eqno(7.34)$$

$$E(2) \cdot P(1) = b \ sinh \mu \eqno(7.35)$$

$$tr M(1) = 1+2 \ cosh \lambda \eqno(7.36)$$

$$tr M(2) = 1+2 \ cosh\mu \eqno(7.37)$$

	As we have seen, the torus is a rather singular case in many
respects.  The case of genus two is more representative of the
general case $g > 1, N \not= 0$.

\noindent 7.2  The Genus Two Universe

	The reduced variables are the three-vectors $E(1), E(2), E(3),
E(4), M(1)$, $M(2), M(3), M(4)$, and the constraints are

$$J \equiv \sum^4_{\mu = 1} \biggl( I - M^{-1}(\mu)\biggr) E(\mu) \approx
0 \eqno(7.38)$$

$$P^a \equiv {1\over 2} \epsilon^{abc} \biggl(
M(1)M^{-1}(2)M^{-1}(1)M(2)M(3)M^{-1}(4)M^{-1}(3)M(4)\biggr)_{cb}
\approx 0 \eqno(7.39)$$

	Given a choice of basis loops $u_1, v_1, u_2, v_2$ [Figure 7.2],
one computes the corresponding $ISO(2,1)$ homotopies as explained
in Sec. 3.

$$\eqalign \rho(u_1) = \pmatrix{
M(1) & (I-M(1)) E(1) - M(1)E(2) + (I-M(1)) OA\cr
 \ & \ \cr
0 & \qquad 1\cr} \eqno(7.40)$$

$$\eqalign \rho(v_1) = \pmatrix{
M^{-1}(2) & (I-M^{-1}(1)-M^{-1}(2)) E(1) \cr
\ & \qquad + (I-M^{-1}(2)) E(2) + (I-M^{-1}(2)) OA\cr\cr \ & \ \cr
0 & \qquad \qquad \qquad 1 \cr} \eqno(7.41)$$

$$\eqalign \rho (u_2) = \pmatrix{
M(3) & (M^{-1}(4) - I - M(3) M^{-1}(4)) E(4)\cr
\  & \qquad + (M^{-1}(3)-I)E(3) + (I - M(3)) OA\cr \ & \ \cr
0 & \qquad \qquad 1\cr} \eqno(7.42)$$

$$\eqalign \rho(v_2) = \pmatrix{
M^{-1}(4) & (M^{-1}(4)-M^{-2}(4)) E(4)\cr
& \qquad -M^{-1}(4)M^{-1}(3)E(3) + (I-M^{-1}(4)) OA\cr \  & \ \cr
0 & \qquad \qquad 1\cr} \eqno(7.43)$$

	The observables are $ISO(2,1)$ invariants taken from these matrices
and any products of them.  A convenient choice of 30 invariants was
proposed by Nelson and Regge [5], who also gave their
brackets and the remaining non-linear relations (the number of
phase space variables for $g = 2$ is $12$).  We are now in a
position where we can compute the $30$ invariants in terms of the
reduced variables $E(\mu )$ and $M(\mu )$.  in [Table 7.1] we give
the 15 loops chosen by Nelson and Regge and the two invariants for
each loop.  We use the condensed notation $P^a(2 {\buildrel \_ \over
{1}} ) = (1/2) \epsilon^{a b c} (M(1)M^{-1}(2))_{cb}$.
In [5], Nelson and Regge
worked with the group $SO(2,2)$ rather than $ISO(2,1)$, which amounts
to having a negative cosmological constant. However the Poincar\'e
limit $\Lambda \to 0$ was carried out explicitly in ref [3]. The invariants
$L_1$ and $L_2$ in this article are equivalent to the traces $a_I \to
a_{XV}$ used in [5] in the limit $\Lambda \to 0$.

     In the expressions from Table 7.1, one identifies the explicit
dependence on time by replacing E(1) by its expression from
gauge-fixing, Eqn. (5.13), as in the example from Sec. 6:

$$L_2 (u_1 v_1) = E(2) \cdot P(2 {\buildrel \_ \over {1}}) + A P^t(2
{\buildrel \_ \over {1}}) - B P^y(2 {\buildrel \_ \over {1}})
- P^x(1) P^x(2 {\buildrel \_ \over{1}}) T \eqno(6.27)$$

     Given the algebraic relations between the variables, one must still
check that the Poisson brackets and constraints are consistent. The
Poisson brackets were shown to be consistent in the general case in
Sec. 6.4. As for the constraints, the $ISO(2,1)$ matrices $u_i$ and
$v_i$ are required to satisfy the cycle condition, specifically:

$$\rho(u_1 v_1 u^{-1}_1 v^{-1}_1 \ u_2 v_2 u_2^{-1} v_2^{-1}) = I
\eqno(7.44)$$

\noindent By computing this product explicitly with the relations
(7.40)-(7.43), one finds

$$\pmatrix{
W & -WJ+(I-W)OA\cr \ & \ \cr
0 & 1\cr} = I \eqno(7.45)$$

     Since the dimension of the reduced phase space for genus two is
$12g-12=12$, the 30 invariants given in the table cannot all be
independent. Indeed, the complete set of relations among them was
proposed by Nelson and Regge. To complete the proof of equivalence
of the formalisms, one must show that these relations are satisfied
by the expressions given in the table above, when the variables
satisfy the constraints (7.38)-(7.39). This is a trivial consequence
of the fact that the matrices $\rho(u_i), \rho(v_i)$ generate a
representation of $\pi_1(\Sigma_g) \to ISO(2,1)$.

\vfill
\eject

\section{Conclusion}

     In (2+1)-dimensional gravity, the physical degrees of freedom
can be parametrized by a representation $\rho: \pi_1(\Sigma_{g, N})
\to ISO(2,1)$. We have shown how elements $\rho(u_i)$ can be written
in terms of the finite set of variables $\{ E(\mu), M(\mu); \mu =
1, \cdots, 2g+N \}$ of the polygon representation. The $ISO(2,1)$
cycle conditions for the representations $\rho$ are equivalent to
the $SO(2,1)$ cycle conditions $P^a \approx 0$ together with the
polygon closure relations $J^a \approx 0$, and the symplectic
structure on the reduced space of $ISO(2,1)$ invariants can be
recovered from the Poisson brackets of the polygon variables. Thus,
the homotopy invariants and the polygon variables represent the
same set of solutions of $(2+1)$-dimensional gravity, and the
phase space of $ISO(2,1)$ invariants is closely related to the
Hamilton-Jacobi phase space of the polygon representation.

     The choice of an internal time in the polygon representation
permits an explicit calculation of the Hamiltonian for any genus.
On the other hand, the virtue of working with the extrinsic slicing
is that one finds expressions which are
similar to those encountered in the ADM formalism for
(3+1)-dimensional gravity. We will pay the price of losing this
similarity, and in exchange hope to push further the quantization
programme for (2+1)-dimensional gravity.

     The method which we used in this article is limited to flat
spacetimes, since the reduced variables $E(\mu), M(\mu)$ have yet to
be generalized to de Sitter gravity (one possible way to do this was
sketched in [30]). Nevertheless, the problem of time is conceptually
similar in (2+1)-dimensional de Sitter gravity and we expect the
conclusion to generalize: The homotopy invariants are most likely
related to the Hamilton-Jacobi phase space for any of
the exactly soluble theories
discussed by Horowitz [31], as well as their supersymmetric
generalizations. A generalization to flat (3+1)-dimensional
gravity (with topological degrees of freedom) could be carried out
following Horowitz and the reduction to global $E - M$ variables, to
define a three-dimensional polygonal cell embedded in $\Real^{3+1}$
[32]. One might also consider the loop space variables of Ashtekar et
al. [33]. The loop variables are related to the homotopy
invariants discussed in this article, but are path-dependent and
uncountable, among other important differences.

    The quantum analogue of the Hamilton-Jacobi formalism is the
Heisenberg picture quantization; observables are constants but depend
explicitly on time. Three highly non-trivial problems come up. One is
the invariance of the wave function with respect to the mapping class
symmetry. So far the only examples of symmetric wave functions for
$(2+1)$-dimensional gravity are for the simplest cases of two
particles scattering on $\Real^2$ [34] and the quantum torus [10].
The second problem is that of ordering the operators in the solution
of the Heisenberg equations of motion
in a way which is consistent with the mapping class group;
this difficulty was emphasized by Carlip [10], who found a consistent
ordering of the relations giving the homotopy invariants for $g = 1$ as
functions of the ADM variables, where the wave functions
transform non-trivially under the modular group. The third is the
so-called ``problem of time'', which can be summarized as follows. In
order to formulate dynamical questions, a choice of ``internal
time'' must be made. The question then is whether the answer to a
given question depends on the choice of time. This problem is
linked to the ordering ambiguities and the problem of consistency with
the mapping class symmetry. Thus, although the classical theory
of (2+1)-dimensional gravity is essentially solved, the status of the
quantum theory is far less clear; we examine it in greater detail in the
following article.

\vfill
\eject

\section{Acknowledgements}

    One of us (HW) would like to express his gratitude to the
Institute for Theoretical Physics (ITP) of the University of
California at Santa Barbara, for their invitation to participate in
the workshop on the small scale structure of spacetime, as well as
the Facultad de Astronom\ii a, Matem\' aticas y F\ii sica of the
University of Cordoba, Argentina, the Facultad de Ciencias Exactas y
Naturales and the Instituto de Astronom\ii a y F\ii sica del Espacio of
the University of Buenos Aires, and the Facultad de Ciencias Exactas
and Instituto de F\ii sica Rosario of the National University of
Rosario, Argentina, for their wonderful hospitality while much of this
work was being completed. We are endebted to Luis Urrutia for enlightening
discussions at the early stages of this work. We also gladly
acknowledge valuable input form e-mail exchanges and conversations
with Steve Carlip and Vince Moncrief.
\

\vfill
\eject

\parskip=6pt
\parindent=0pc

\section{References}

[1]	Ed Witten, Nucl. Phys. B311 (1988) 46

[2]	J. Soda and Y. Yamanaka, Mod. Phys. Letters A 6 (1991), 303

$\ \ \ $	A. Ashtekar and J. D. Romano, Phys. Lett. B229 (1989), 56

[3]	L. F. Urrutia and F. Zertuche, Class. Quant. Grav. 9 (1992), 641

[4]	J. E. Nelson, T. Regge, F. Zertuche, Nucl. Phys. B339 (1990), 516

[5]	J. E. Nelson and T. Regge, Commun. Math. Phys. 141 (1991), 211

[6]	J. E. Nelson and T. Regge, Phys. Lett. B272 (1991), 213

[7]	V. Moncrief, Ann. Phys. (N.Y.) 167 (1986), 118

$\ \ \ $	V. Moncrief, J. Math. Phys. 30 (1989), 2297

[8]	R. A. Matzner and L. C. Shepley, Classical Mechanics, Prentice
Hall (Englewood Cliffs, N. J., 1991).

[9]	R. Arnowitt, S. Deser and C. W. Misner, Gravitation, and
Introduction to Current Research, [L, Witten, ed.]; Wiley N.Y. (1962)

$\ \ \ $	J. York, Phys. Rev. Lett. 28 (1972), 1082

[10]	S. Carlip, Phys. Rev. D45 (1992), 3584

$\ \ \ \ \ $	S. Carlip, The modular group, operator ordering and time in 2+1
dimensional gravity,  Univ. of California Preprint UCD-92-23
(gr-qc/9209011)

[11]	A. Anderson: ``Unitary equivalence of the metric and holonomy
formulations of 2+1 dimensional quantum gravity on a torus",
Mc Gill University preprint 92-19, Imperial College preprint
TP/92-93/02  (gr-qc/921007).

$\ \ \ \ \ $	A. Anderson: ``Quantum canonical transformations:  Physical
equivalence of quantum theories", Imperial College preprint
TP/92-93/20  (hep-th/9302062).

[12]	H. Waelbroeck, Nucl. Phys. B364 (1991), 475

[13]	S. Carlip, private communication

[14]	J. D. Brown, Lower-dimensional Gravity  (World Scientific, 1988)

[15]	A. Hosoya and K. Nakao, Class. Quant. Grav. 7 (1990), 63

$\ \ \ \ \ $	(see also ref. 3)

[16]	A. Achucarro and P. Townsend, Phys. Lett. B180 (1986), 89

[17] 	J. E. Nelson, T. Regge, Nucl. Phys. B328 (1989), 190

[18]	L.F. Urrutia, H. Waelbroeck, F. Zertuche, Mod. Phys. Lett.
A7 (1992), 2715

[19]	P.A.M. Dirac, Lectures on Quantum Mechanics, (New York:
Belfer Graduate School of Science Monographs Series, 1964).

[20]	H. Waelbroeck, Class. Quant. Grav. 7 (1990), 751

$\ \ \ \ \ $	H. Waelbroeck, Phys. Rev. Lett. 64 (1990), 2222

[21]	H. Waelbroeck, The Polygon Representation of Flat
Three-dimensional Spacetimes,  Rev. Mex. Fis. 6 No. 39 (1993), 831

[22]	G. Mess, Lorentz spacetime of constant curvature, Institut
des Hautes Etudes Scientifiques Preprint IHES/M/90/28 (1990)

[23]	S. Carlip, Class. Quantum Grav.8 (1991), 5

$\ \ \ \ \ $	S. Carlip, Nucl. Phys. B324 (1989), 106

[24]	H. Poincar\' e, Acta Mathematica 1 (1882), 1

[25]	B. S. DeWitt, Phys. Rev. 160 (1967), 1113

[26]	K. Kuchar, Time and interpretations of quantum gravity,
Proceedings of the fourth Canadian Conference on General
Relativity and Relativistic Astrophysics, eds. G. Kunstatter,
D. Vincent and J. Wiliams (World Sci., Singapore, 1992)

[27]	S. Deser, R. Jackiw and G. 't Hooft, Ann. Phys. (N. Y.) 152
(1984), 220

[28]	A. Guth, private communication

[29]	S. Deser, R. Jackiw and G. 't Hooft, Phys. Rev. Lett. 68
(1992), 267

[30]	H. Waelbroeck, L. Urrutia and F. Zertuche, in Proceedings
of the Sixth Marcel Grossman Meeting  (World Scientific, 1991)

[31]	G. T. Horowitz, Commun. Math. Phys. 125 (1989), 417

[32]	H. Waelbroeck and J. A. Zapata, A Hamiltonian Lattice
Formulation of Topological Gravity, to appear in Class. Quantum
Grav. (1994)

$\ \ \ \ \ $ H. Waelbroeck, $B \wedge F$ Theory and Flat Spacetimes,
Preprint ICN-UNAM-93-12 (gr-qc/9311033)

[33]	A. Ashtekar, Phys. Rev. Lett. 57 (1987), 2244

$\ \ \ \ \ $	A. Ashtekar, Phys. Rev. D36 (1987), 1587

$\ \ \ \ \ $	T. Jacobson and L. Smolin, Nonperturbative Quantum
Cosmologies, Nucl. Phys. B299 (1988), 295

$\ \ \ \ \ $	A. Ashtekar, Canonical quantum gravity,
Proceedings of the 1990 Banff Summer School on Gravitation,
ed. by R. Mann (World Sci., Singapore, 1990)

$\ \ \ \ \ $	C. Rovelli, Ashtekar formulation of General
Relativity and Loop Space Non-perturbative Quantum Gravity:
 A Report,  Class. Quant. Grav. 8 (1991)

[34]	G. 't Hooft, Commun. Math. Phys. 117 (1988), 685

$\ \ \ \ \ $	S. Deser and R. Jackiw, Commun. Math. Phys. 118
(1988), 495

\vfill
\eject

\section{Figure Captions}

Figure 2.1  A basis for the homotopy group of $\Sigma_g$
is represented.  Note that each $u_i$ $(i=1,...,g)$ intersects
only the corresponding $v_i$, at only one point.  These
loops are not independent: the last one ($v_g$) can be
deformed into a combination of the other $2g-1$.

\

Figure 3.1  The torus (initial surface) is represented by a
parallelogram with opposite sides identified.  At each corner,
one has an image of the timelike line $N$.  The evolution of
this torus is obtained by sliding each corner along the
corresponding line.

\

Figure 4.1  One handle of a genus $g$ surface is represented by
four edges of a polygon, with identifications
$E(1) \sim -M^{-1}(1)E(1), E(2) \sim -M^{-1}(2)E(2)$.  The loops
$u_1$ and $v_1$ cross only at $O$, which can be chosen arbitrarily
on $\Sigma_g$.

\

Figure 7.1  The path $u v u^{-1} v^{-1}$ on the torus is contractible.
To see this, inflate the area which is squeezed between
the double lines until it becomes a single line surrounding
the four corners.  Since these four corners are identified,
this line is a small circle around the ``vertex": the point of
$T^2$ which maps onto the four corners.  This small circle is
contractible, since there is no topological obstruction at
the vertex.

\

Figure 7.2  The genus two spacetime is represented by an
octagon with the identifications $E(\mu) \sim -M^{-1}(\mu)E(\mu)$,
$\mu = 1, 2, 3, 4$.  The four loops $u_1, v_1, u_2, v_2$,
which intersect only at $O$, form a basis of the homotopy
group based at $O$.

\

Table 7.1  The 30 homotopy invariants proposed by Nelson and
Regge are given in terms of the polygon variables.

\end